\documentclass[pasp]{emulateapj}
\slugcomment{}

\usepackage{graphicx} 
\usepackage{amssymb}  
\usepackage{multirow} 

\newcommand{\nin}{\noindent}
\newcommand{\unit}[1]{\textrm{ #1}}

\newcommand{\rproc}{{\em r}-process}

\newcommand{\euii}{Eu~{\sc ii}}
\newcommand{\fei}{Fe~{\sc i}}

\newcommand{\loggf}{$\log{\textrm{\em gf}}$}
\newcommand{\gamsix}{$\Gamma_{6}$}

\newcommand{\kms}{km s$^{-1}$}
\newcommand{\teff}{$T_{\rm eff}$}
\newcommand{\mteff}{T_{\rm eff}}
\newcommand{\logg}{$\log{g}$}
\newcommand{\mlogg}{\log{g}}
\newcommand{\feh}{[Fe/H]}
\newcommand{\monh}{[M/H]}
\newcommand{\euh}{[Eu/H]}
\newcommand{\chisq}{$\chi^2$}
\newcommand{\rchisq}{reduced $\sqrt{\chi^2}$}
\newcommand{\sA}{\,\AA}

\newcommand{\nstarsn}{41}
\newcommand{\rchisqtot}{2.78}  
\newcommand{\rchisqsm}{1.57}   
\newcommand{\rchisqlit}{1.93}  
\newcommand{\linedatafootnote}{All line data except \gamsix\ from Kurucz databases via VALD (unless otherwise noted).}
\newcommand{\vanderwaalsfootnote}{Van der Waals parameters where available from \citealt{barklem_2000} via VALD (unless otherwise noted).}
\newcommand{\moleculesfootnote}{All molecular line data (except \gamsix, from \chisq\ minimization) from Kurucz web site:\ \texttt{http://kurucz.harvard.edu/LINELISTS/LINESMOL/}.}
\newcommand{\moorefootnote}{Identification from \citealt{moore_1966}, \loggf\ and \gamsix\ from \chisq\ minimization.}
\newcommand{\kuruczonlinefootnote}{Identification and parameters from Kurucz line lists hosted by the University of Hannover:\newline \texttt{http://www.pmp.uni-hannover.de/cgi-bin/ssi/test/kurucz/sekur.html}.}

\received{22 April 2009}
\accepted{19 May 2009}
\shortauthors{K.\,M.\,G.\,Peek}
\shorttitle{Measuring Europium Abundances}
 
\begin{document}

\title{A New Technique for Determining Europium Abundances in Solar-Metallicity Stars\altaffilmark{1}}
\author{Kathryn M.\,G.~Peek\altaffilmark{2}}
\email{kpeek@astron.berkeley.edu}

\altaffiltext{1}{The data presented herein were obtained at the W.\,M.\ Keck Observatory, which is operated as a scientific partnership among the California Institute of Technology, the University of California, and the National Aeronautics and Space Administration. The Observatory was made possible by the generous financial support of the W.\,M.\ Keck Foundation.}
\altaffiltext{2}{Department of Astronomy, University of California, 
Berkeley, CA USA 94720-3411}

\begin{abstract}
We present a new technique for measuring the abundance
of europium, a representative \rproc\ element, 
in solar-metallicity stars.
Our algorithm compares LTE synthetic spectra with
high-resolution observational spectra 
using a \chisq-minimization routine.
The analysis is fully automated, and therefore
allows consistent measurement of blended lines 
even across very large stellar samples.
We compare our results with
literature europium abundance measurements and
find them to be consistent; we also find our method 
generates smaller errors.
\end{abstract}


\section{Introduction}
\label{intro}
Every atom heavier than lithium has been processed by stars.  
Elements of the $\alpha$ process, iron peak, and 
{\em r} process have different formation sites, 
and therefore, understanding the distribution of 
these elements in nearby stars can lead to a 
better understanding of the Galaxy's 
chemical enrichment history.
The \rproc\ site is the least well understood.
Europium is our choice for an \rproc\ investigation
for two reasons:\ 96\% of Galactic europium is 
formed through the \rproc\ \citep{burris_2000}, 
and it has several strong lines in the 
visible portion of the electromagnetic spectrum.
In order to provide insight into the Galactic
enrichment history, however, europium measurements
in very large stellar samples will be needed.

The study of \rproc\ elements is well established in metal-poor 
stars \citep[e.g.,][]{sneden_2008,lai_2008,frebel_2007,
simmerer_2004,johnson_2001},
where \rproc\ abundances probe early Galactic history.  
A handful of studies have measured europium in 
solar-metallicity stars 
\citep[e.g.,][]{reddy_2006,bensby_2005,koch_2002,woolf_1995}, 
but have done so in samples of 50--200 stars.
We aim to extend such work to a substantial sample of 
1000 stars at solar metallicity, using spectra
collected by the California and Caltech Planet Search
(CCPS).
Such a large sample will require automated analysis.

Our goal in this paper is to establish a new, 
automated abundance fitting method based on Spectroscopy Made Easy
\citep[SME;][]{valenti_1996}, an LTE spectral synthesis code 
discussed in further detail in \S\,\ref{abund}. 
The analysis builds on the framework of the 
\citealt{valenti_2005} (hereafter VF05) Spectroscopic 
Properties of Cool Stars (SPOCS) catalog.
Our method will yield consistent results across large stellar 
samples, generating smaller errors than previous analyses.

We begin here with \nstarsn\ stars that
have been examined in the literature, and we include 
three europium lines:\ 4129\sA, 4205\sA, and 6645\sA.
Our stellar observations are detailed in \S\,\ref{obs}.
We describe the details of our europium 
abundance measurement algorithm in \S\,\ref{abund}, 
compare our values with existing europium literature
in \S\,\ref{comp},
and summarize the results of the study in 
\S\,\ref{summary}.

\section{Observations}
\label{obs}

The spectra of the \nstarsn\ stars included in this study were taken with the HIRES echelle spectrograph \citep{vogt_1994} on the 10-m Keck I telescope in Hawaii.  
The spectra date from January 2004 to September 2008 and have resolution $R \sim 50\,000$ and signal-to-noise ratio (S/N) of $\sim$\,$160$ near \mbox{4200\sA}, where the two strongest \euii\ lines included in this study are located.  
The spectra have the same resolution but S/N~$\sim 340$ at \mbox{6645\sA}, where a third, weaker \euii\ line is located. 

The spectra were originally obtained by the CCPS with the intention of detecting exoplanets. 
The CCPS uses the same spectrometer alignment each night and employs the HIRES exposure meter \citep{kibrick_2006}, so the stellar observations are extremely consistent, even across years of data collection. 
For a more complete description of the CCPS and its goals, see \citealt{marcy_2008}.
It should be noted that the iodine cell Doppler technique \citep{marcy_1992} imprints molecular iodine lines only between the wavelengths of 5000 and 6400\sA, leaving the regions of interest for this study iodine free.

\subsection{Stellar Sample}
\label{stellsamp}
For this study we choose to focus on a subset of 
stars that have been analyzed in other abundance studies.  
We compare our measurements with those from 
\citealt{woolf_1995,simmerer_2004,bensby_2005}; and 
\citealt{delpeloso_2005a}. We select CCPS target stars that 
have temperature, gravity, mass, and metallicity measurements 
in the SPOCS catalog (VF05). 

Our initial stellar sample consisted of 44 objects that were 
members both of the aforementioned studies and the SPOCS catalog, 
and that were observed by the CCPS on the Keck I telescope.
Three stars that otherwise fit our criteria, but for which we
could only measure the europium abundance in one line, were 
removed from the sample.
They were HD\,64090, HD\,188510, and HIP\,89215. 
While stellar europium abundances may be successfully 
measured from a single line, 
the goal of this study is to establish the robustness 
of our method, 
and so we deem it necessary to determine the 
europium abundance in at least two lines to 
include a star in our analysis here.
We describe our criteria for rejecting a fit in 
\S\,\ref{stellar_abund}.

The \nstarsn\ stars included in this study have 
$3.4 < V < 10.0$,
$0.50 < B-V < 0.92$, and
$5 < d < 75\unit{pc}$ ({\em Hipparcos}; \citealt{hipparcos}).
VF05 determine stellar properties using SME,
so it is reasonable to adopt their values for our SME europium
analysis.
Based on the VF05 analysis, our stars have
metallicity $-1.4 < \textrm{\monh} < 0.4$,
effective temperature $4940 < \mteff < 6230$, and
gravity $3.8 < \mlogg < 4.8$.
It should be noted that throughout this study we use the 
\monh\ parameter for a star's metallicity, rather than 
the iron abundance \feh. 
Our \monh\ value is taken from VF05, where it 
is an independent model parameter that adjusts the
relative abundance of all metals together.
It is not an average of individual metal 
abundances.
The full list of stellar properties, from the {\em Hipparcos} and
SPOCS catalogs, appears in Table \ref{starsinfo_table}.

\begin{deluxetable*}{rrrrrrrrrc}
\tablecaption{Stellar Data.\label{starsinfo_table}}
\tablewidth{0pt}
\tablehead{
  \multicolumn{3}{c}{{Identification}}
  & \colhead{$V$\tablenotemark{a}}
  & \colhead{$B-V$\tablenotemark{a}}
  & \colhead{$d$\tablenotemark{a}}
  & \colhead{\teff\tablenotemark{b}}
  & \multirow{2}{*}{\monh\tablenotemark{b}}
  & \colhead{$\log{g}$\tablenotemark{b}}
  & \multirow{2}{*}{Ref.\tablenotemark{c}} \\
  \colhead{HD}
  & \colhead{HR}
  & \colhead{HIP}
  & \colhead{(mag)}
  & \colhead{(mag)}
  & \colhead{(pc)}
  & \colhead{(K)}
  & 
  & \colhead{(cgs)}
  & 
}
\startdata 
    3795 &     173 &    3185 &  6.14 &  0.72 &  28.6 &  5369 & $ -0.41$ &  4.16 &       1 \\
    4614 &     219 &    3821 &  3.46 &  0.59 &   6.0 &  5941 & $ -0.17$ &  4.44 &       4 \\
    6734 &  \ldots &    5315 &  6.44 &  0.85 &  46.4 &  5067 & $ -0.28$ &  3.81 &       1 \\
    9562 &     448 &    7276 &  5.75 &  0.64 &  29.7 &  5939 & $  0.19$ &  4.13 & 1, 2, 4 \\
    9826 &     458 &    7513 &  4.10 &  0.54 &  13.5 &  6213 & $  0.12$ &  4.25 &       4 \\
   14412 &     683 &   10798 &  6.33 &  0.72 &  12.7 &  5374 & $ -0.45$ &  4.69 &       1 \\
   15335 &     720 &   11548 &  5.89 &  0.59 &  30.8 &  5891 & $ -0.20$ &  4.07 &       4 \\
   16397 &  \ldots &   12306 &  7.36 &  0.58 &  35.9 &  5788 & $ -0.35$ &  4.50 &       1 \\
   22879 &  \ldots &   17147 &  6.68 &  0.55 &  24.3 &  5688 & $ -0.76$ &  4.41 & 1, 2, 4 \\
   23249 &    1136 &   17378 &  3.52 &  0.92 &   9.0 &  5095 & $  0.03$ &  3.98 &       1 \\
   23439 &  \ldots &   17666 &  7.67 &  0.80 &  24.5 &  5070 & $ -0.73$ &  4.71 &       3 \\
   30649 &  \ldots &   22596 &  6.94 &  0.59 &  29.9 &  5778 & $ -0.33$ &  4.44 &       4 \\
   34411 &    1729 &   24813 &  4.69 &  0.63 &  12.6 &  5911 & $  0.09$ &  4.37 &       4 \\
   43947 &  \ldots &   30067 &  6.61 &  0.56 &  27.5 &  5933 & $ -0.28$ &  4.37 &       2 \\
   45184 &    2318 &   30503 &  6.37 &  0.63 &  22.0 &  5810 & $  0.03$ &  4.37 &       1 \\
   48938 &    2493 &   32322 &  6.43 &  0.55 &  26.6 &  5937 & $ -0.39$ &  4.31 &       4 \\
   84737 &    3881 &   48113 &  5.08 &  0.62 &  18.4 &  5960 & $  0.14$ &  4.24 &       4 \\
   86728 &    3951 &   49081 &  5.37 &  0.68 &  14.9 &  5700 & $  0.11$ &  4.29 &       4 \\
  102365 &    4523 &   57443 &  4.89 &  0.66 &   9.2 &  5630 & $ -0.26$ &  4.57 &       2 \\
  \ldots &  \ldots &   57450 &  9.91 &  0.58 &  73.5 &  5272 & $ -1.42$ &  4.30 &       3 \\
  103095 &    4550 &   57939 &  6.42 &  0.75 &   9.2 &  4950 & $ -1.16$ &  4.65 &       3 \\
  109358 &    4785 &   61317 &  4.24 &  0.59 &   8.4 &  5930 & $ -0.10$ &  4.44 &       4 \\
  115617 &    5019 &   64924 &  4.74 &  0.71 &   8.5 &  5571 & $  0.09$ &  4.47 &       4 \\
  131117 &    5542 &   72772 &  6.30 &  0.61 &  40.0 &  5973 & $  0.10$ &  4.06 &       2 \\
  144585 &  \ldots &   78955 &  6.32 &  0.66 &  28.9 &  5854 & $  0.25$ &  4.33 &       1 \\
  156365 &  \ldots &   84636 &  6.59 &  0.65 &  47.2 &  5856 & $  0.24$ &  4.09 &       1 \\
  157214 &    6458 &   84862 &  5.38 &  0.62 &  14.4 &  5697 & $ -0.15$ &  4.50 &       4 \\
  157347 &    6465 &   85042 &  6.28 &  0.68 &  19.5 &  5714 & $  0.03$ &  4.50 &       1 \\
  166435 &  \ldots &   88945 &  6.84 &  0.63 &  25.2 &  5843 & $  0.01$ &  4.44 &       1 \\
  169830 &    6907 &   90485 &  5.90 &  0.52 &  36.3 &  6221 & $  0.08$ &  4.06 &       1 \\
  172051 &    6998 &   91438 &  5.85 &  0.67 &  13.0 &  5564 & $ -0.24$ &  4.50 &       1 \\
  176377 &  \ldots &   93185 &  6.80 &  0.61 &  23.4 &  5788 & $ -0.23$ &  4.40 &       1 \\
  179949 &    7291 &   94645 &  6.25 &  0.55 &  27.0 &  6168 & $  0.11$ &  4.34 &       1 \\
  182572 &    7373 &   95447 &  5.17 &  0.76 &  15.1 &  5656 & $  0.36$ &  4.32 &       1 \\
  190360 &    7670 &   98767 &  5.73 &  0.75 &  15.9 &  5552 & $  0.19$ &  4.38 &       1 \\
  193901 &  \ldots &  100568 &  8.65 &  0.55 &  43.7 &  5408 & $ -1.19$ &  4.14 &       3 \\
  199960 &    8041 &  103682 &  6.21 &  0.64 &  26.5 &  5962 & $  0.24$ &  4.31 &       1 \\
  210277 &  \ldots &  109378 &  6.54 &  0.77 &  21.3 &  5555 & $  0.20$ &  4.49 &       1 \\
  217107 &    8734 &  113421 &  6.17 &  0.74 &  19.7 &  5704 & $  0.27$ &  4.54 &       1 \\
  222368 &    8969 &  116771 &  4.13 &  0.51 &  13.8 &  6204 & $ -0.08$ &  4.18 &       4 \\
  224383 &  \ldots &  118115 &  7.89 &  0.64 &  47.7 &  5754 & $ -0.06$ &  4.31 &       1 \\
\enddata
\tablenotetext{a}{$V$-magnitude, color index, and parallax-based distance 
from the {\em Hipparcos} catalogue \citep{hipparcos}.}
\tablenotetext{b}{Stellar parameters previously published in VF05.}
\tablenotetext{c}{Star included for comparison to the following 
works:~1---\citealt{bensby_2005};
2---\citealt{delpeloso_2005a};
3---\citealt{simmerer_2004};
4---\citealt{woolf_1995}}
\end{deluxetable*}

\subsection{Co-adding Data}
\label{data}
The nature of the radial velocity planet search dictates that most stars 
will have multiple, and in some cases, dozens of, observations.  
To take advantage of the multiple exposures, we carefully 
co-add the reduced echelle spectra where possible. 

The co-adding procedure is as follows:\ a 2000-pixel region 
(approximately half the full order width) near 
the middle of each order is cross-correlated, order by order, 
with one observation arbitrarily designated as standard. 
The pixel shifts are then examined and a linear trend as a function of 
order number is fit to the pixel shifts.  
For any order whose pixel shift falls more than 0.4 pixels from the linear 
trend, the value predicted by the linear trend is substituted.  
This method corrects outlying pixel shift values, which often proved to
be one of a handful of problematic orders where the echelle blaze
function shape created a false cross-correlation peak.

Each spectral order is adjusted by its appropriate fractional pixel
shift before all the newly aligned spectra are added together.
In order to accurately add spectra that have been shifted by 
non-integer pixel amounts, we use a linear interpolation between 
pixel values to artificially (and temporarily) increase 
the sampling density by a factor of 20.
After the co-adding, the resultant high-S/N 
multi-observation spectrum is sampled back down to its 
original spacing. 

\begin{figure}  
\includegraphics[width=\columnwidth]{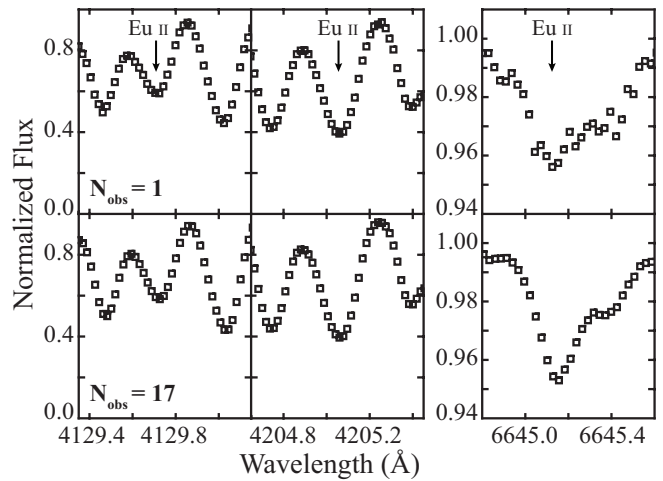}
\figcaption[Co-added Stellar Spectra.]{The three \euii\ lines considered in this paper, 
  plotted for two different stars to demonstrate the advantage to be gained
  from co-adding multiple observations of the same star. 
  Note that the rightmost plots have a different ordinate axis scaling than 
  the other four panels.
  The top three panels, showing HD\,156365, include 1 spectrum, while the bottom 
  three panels, showing HD\,144585, include 17. 
  The two stars have approximately the same metallicity and
  effective temperature. 
  The advantage from co-adding is most profound
  in the weak 6645-\AA\ line.
  \label{coadding_works}}
\end{figure}

The number of observations used per star is recorded in the $N_{obs}$ 
column of Table \ref{starsvals_table}. 
The co-adding proved particularly beneficial when fitting the relatively 
weak \euii\ line at 6645\sA.  
Two sample stellar spectra, from a star with 1 observation
and from a star with 17 observations, are given in Figure 
\ref{coadding_works} as evidence of the S/N advantages 
of this procedure.

\begin{deluxetable}{rrrrrrr}
\tablecaption{Stellar abundance values.\label{starsvals_table}}
\normalsize
\tablehead{
  \multirow{2}{*}{Name\tablenotemark{a}} 
  & \multirow{2}{*}{$N_{obs}$}
  & \colhead{4129 \AA}
  & \colhead{4205 \AA}
  & \colhead{6645 \AA}
  & \colhead{Weighted} \\
 & 
  & \colhead{[Eu/H]\tablenotemark{b}}
  & \colhead{[Eu/H]\tablenotemark{b}}
  & \colhead{[Eu/H]\tablenotemark{b}}
  & \colhead{Average\tablenotemark{c}}
}
 
\startdata
        3795 &    2 &  $0.07$ &  $0.06$ &      $0.13$ &$0.07 \pm    0.03$ \\
        4614 &    2 & $-0.17$ & $-0.19$ &     $-0.24$ &$-0.18 \pm   0.02$ \\
        6734 &    6 & $-0.02$ & $-0.06$ &      $0.01$ &$-0.03 \pm   0.03$ \\
        9562 &    3 &  $0.11$ &  $0.14$ &      $0.08$ &$0.12 \pm    0.02$ \\
        9826 &    1 &  $0.10$ &  $0.13$ &      $0.09$ &$0.12 \pm    0.02$ \\
       14412 &   70 & $-0.28$ & $-0.15$ &     $-0.16$ &$-0.24 \pm   0.06$ \\
       15335 &    2 & $-0.13$ & $-0.10$ &     $-0.08$ &$-0.12 \pm   0.03$ \\
       16397 &    1 & $-0.21$ & $-0.20$ &     $-0.22$ &$-0.21 \pm   0.02$ \\
       22879 &   27 & $-0.55$ & $-0.50$ &     $-0.57$ &$-0.54 \pm   0.03$ \\
       23249 &   34 &  $0.05$ &  $0.26$ &      $0.24$ &$0.17 \pm    0.11$ \\
       23439 &   24 & $-0.43$ & $-0.36$ &     $-0.47$ &$-0.38 \pm   0.03$ \\
       30649 &    3 & $-0.13$ & $-0.11$ &     $-0.13$ &$-0.12 \pm   0.02$ \\
       34411 &   70 &  $0.12$ &  $0.13$ &      $0.11$ &$0.13 \pm    0.02$ \\
       43947 &    6 & $-0.21$ & $-0.19$ &     $-0.21$ &$-0.20 \pm   0.02$ \\
       45184 &   95 &  $0.01$ & $-0.03$ &      $0.02$ &$0.00 \pm    0.02$ \\
       48938 &    2 & $-0.30$ & $-0.28$ &     $-0.41$ &$-0.30 \pm   0.03$ \\
       84737 &    7 &  $0.14$ &  $0.14$ &      $0.14$ &$0.14 \pm    0.02$ \\
       86728 &   46 &  $0.06$ &  $0.11$ &      $0.10$ &$0.09 \pm    0.03$ \\
      102365 &   12 & $-0.13$ & $-0.08$ &     $-0.14$ &$-0.11 \pm   0.03$ \\
    HIP57450 &    1 & $-1.10$ & $-1.03$ &$\leq -0.95$ &$-1.06 \pm   0.04$ \\
      103095 &    9 &  \ldots & $-0.37$ &     $-0.38$ &$-0.37 \pm   0.02$ \\
      109358 &   47 & $-0.12$ & $-0.16$ &     $-0.12$ &$-0.14 \pm   0.03$ \\
      115617 &  165 &  $0.05$ &  $0.02$ &      $0.09$ &$0.04 \pm    0.03$ \\
      131117 &    2 &  $0.10$ &  $0.09$ &      $0.12$ &$0.10 \pm    0.02$ \\
      144585 &   17 &  $0.22$ &  $0.25$ &      $0.22$ &$0.23 \pm    0.03$ \\
      156365 &    1 &  $0.11$ &  $0.20$ &      $0.14$ &$0.13 \pm    0.04$ \\
      157214 &   24 &  $0.05$ &  $0.06$ &      $0.07$ &$0.06 \pm    0.02$ \\
      157347 &   77 &  $0.09$ &  $0.12$ &      $0.13$ &$0.10 \pm    0.02$ \\
      166435 &    1 & $-0.05$ & $-0.17$ &      $0.04$ &$-0.07 \pm   0.06$ \\
      169830 &    8 &  $0.02$ &  $0.05$ &      $0.05$ &$0.03 \pm    0.02$ \\
      172051 &   37 & $-0.19$ & $-0.15$ &     $-0.12$ &$-0.17 \pm   0.03$ \\
      176377 &   56 & $-0.20$ & $-0.21$ &     $-0.23$ &$-0.20 \pm   0.02$ \\
      179949 &    1 &  $0.05$ &  $0.02$ &      $0.05$ &$0.04 \pm    0.03$ \\
      182572 &   56 &  $0.25$ &  $0.41$ &      $0.36$ &$0.31 \pm    0.08$ \\
      190360 &   73 &  $0.17$ &  $0.29$ &      $0.30$ &$0.24 \pm    0.07$ \\
      193901 &    2 & $-0.92$ & $-0.91$ &     $-0.98$ &$-0.91 \pm   0.02$ \\
      199960 &    3 &  $0.14$ &  $0.19$ &      $0.19$ &$0.16 \pm    0.03$ \\
      210277 &   74 &  $0.20$ &  $0.34$ &      $0.32$ &$0.27 \pm    0.07$ \\
      217107 &   37 &  $0.20$ &  $0.43$ &      $0.33$ &$0.29 \pm    0.11$ \\
      222368 &    2 &  $0.08$ &  $0.09$ &      $0.08$ &$0.08 \pm    0.02$ \\
      224383 &    2 &  $0.00$ & $-0.01$ &      $0.04$ &$0.00 \pm    0.02$ \\
       Vesta &    3 & $-0.04$ &  $0.00$ &      $0.00$ &$-0.03 \pm   0.03$ \\
\enddata
\tablenotetext{a}{All names are HD numbers unless otherwise indicated.}
\tablenotetext{b}{For a given star, [Eu/H] = $\log{\epsilon\left(\textrm{Eu}\right)} - \log{\epsilon\left(\textrm{Eu}\right)_{\odot}}$, where \mbox{$\log{\epsilon\left(X\right)} = \log_{10}{\left(N_{X}/N_{H}\right)}$.}}
\tablenotetext{c}{The weight of each line in the average is based on 50 Monte Carlo trials in each \euii~line. We have adopted an error floor of $0.02\unit{dex}$, added in quadrature to the errors determined by our Monte Carlo procedure. See \S\ref{errors} for a more complete description of the weighted average and associated uncertainty.}
\end{deluxetable}

\section{Abundance Measurements}
\label{abund}

We use the SME suite of routines for our spectral synthesis, both for fine-tuning line lists based on the solar spectrum (\S\,\ref{linelists}) and for measuring europium in each star (\S\,\ref{stellar_abund}).
SME is an LTE spectral synthesis code based on the \citealt{kurucz_1992} grid of stellar atmosphere models. 
In brief, to produce a synthetic spectrum, SME interpolates between the atmosphere models, calculates the continuous opacity, computes the radiative transfer, and then applies line broadening, which is governed by macroturbulence, stellar rotation, and instrumental profile. 
Consult \citealt{valenti_1996} and VF05 for a more in-depth description of SME's inner workings.

Throughout this study we use SME only to compute synthetic spectra.
All fitting is done in specialized routines of our own design, external to SME.

In this section we outline our technique for measuring europium abundances in the set of \nstarsn\ stars included in this work. 
In broad strokes, we first determine the atomic parameters of our spectral lines by fitting the solar spectrum (\S\,\ref{linelists}). 
Then, we use that line list to measure the europium abundance in three transitions (4129\sA, 4205\sA, and 6645\sA); a weighted average of the three transitions determines a star's final europium value (\S\,\ref{stellar_abund}). Finally, we estimate our uncertainties by adding artificial noise to our data in a series of Monte Carlo trials (\S\,\ref{errors}).
We also here include notes on individual lines (\S\,\ref{individual_lines}).

\subsection{Line Lists}
\label{linelists}
We use relatively broad spectral segments in our europium 
analysis.
The regions centered on the \euii\ 4129\sA\ and 
6645\sA\ lines are 5\sA\ wide, and the region centered
on the \euii\ 4205\sA\ line is 8\sA\ wide.
We find it necessary to use such broad spectral segments
in order to fit a robust and consistent continuum in 
the crowded blue regions.

Line lists are initially drawn from the Vienna Astrophysics Line 
Database (VALD; \citealt{piskunov_1995,kupka_1999}). 
The VALD line lists, in the regions surrounding all three 
\euii\ transitions, make extensive use of Kurucz line lists. 

We apply the original VALD line list to an observed solar spectrum in
order to determine the list's completeness and to adjust line parameters
as needed.
In the blue, we use the disk-center solar spectrum from 
\citealt{wallace_1998} with the following global parameters: 
\teff~$ = 5770\unit{K}$, 
$\log{g} = 4.44$, 
\monh~$ = 0$, 
microturbulence $v_{mic} = 1.0\unit{\kms}$, 
macroturbulence $v_{mac} = 3.6\unit{\kms}$,
rotational velocity $v\sin{i} = 0\unit{\kms}$, 
and radial velocity $v_{rad} = 0\unit{\kms}$. 
These are the same solar parameters adopted in VF05.
In the red, we find the \citealt{wallace_1998} solar atlas 
to have insufficient S/N to accurately determine
the atomic parameters.
At 6645\sA, therefore, we instead compare our 
line list to the disk-integrated NSO solar spectrum 
\citep{kurucz_1984}, 
adjusting $v\sin{i}$ to $1.6\unit{\kms}$ because
the full solar disk has more substantial rotational 
broadening.

We find that adjustments to the oscillator strength 
(\loggf) and van der Waals broadening (\gamsix) parameters 
are required for the strongest lines in a given 
wavelength segment, even far from the \euii\ line of 
interest.
For example, the \fei\ line at 4202\sA\ has an equivalent 
width $W=326\unit{m\AA}$ in the Sun 
and significantly affects 
the continuum of the 4205-\AA\ region.
The \loggf\ parameter controls the line depth while the
\gamsix\ parameter controls the line shape, so in general
the two parameters are orthogonal.

We used the Kurucz \loggf\ values provided 
by VALD where possible, but adjustments
were necessary where line depths were poorly fit.
For \gamsix, VALD returns the \citealt{barklem_2000} 
parameters for beryllium through barium ($Z$ of 4--56),
but has no \gamsix\ values above $Z=56$.
We therefore find it necessary to fit \gamsix\ in 
species heavier than barium and in deep features
not fit well by VALD values.
We take particular care to determine the appropriate 
\loggf\ and \gamsix\ parameters for all lines 
adjacent to the \euii\ line of interest.  
We find the best value for these parameters with the SME 
synthesizer by performing a \chisq\ minimization against 
the solar spectrum on each parameter.

\begin{figure}  
\includegraphics[width=\columnwidth]{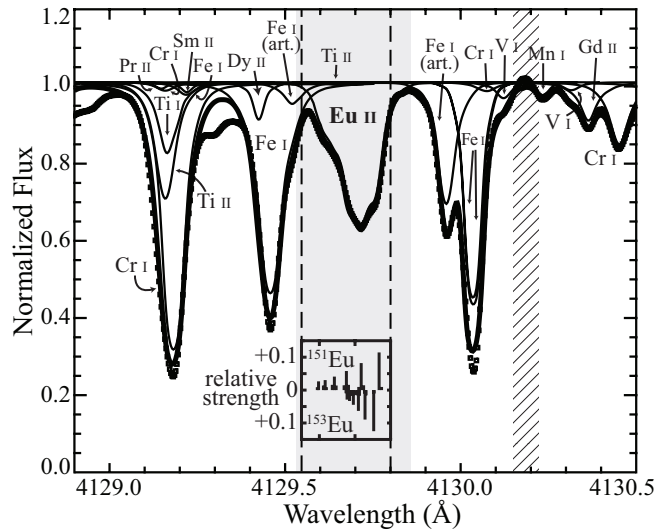}
\figcaption[Eu 4129\sA\ in NSO.]{The 4129-\AA\ \euii\ line in the solar spectrum. 
  Individual lines are annotated and are listed in Table \ref{4129_nso_table}.
  The hyperfine components (see \S\,\ref{linelists}) appear in the inset plot, 
  which is aligned with the wavelength scale of the main plot.
  The relative strengths of the 32 hyperfine components \citep{ivans_2006}
  are plotted on a linear scale in the inset; the top half of the inset 
  contains the components from the $^{151}$Eu isotope while the
  $^{153}$Eu isotope components appear on the bottom.
  The gray box indicates the portion of the spectrum used to calculate 
  \chisq\ during the abundance fitting step (see \S\,\ref{stellar_abund}).
  The cross-hatched region indicates a portion of the spectrum 
  used to fit a continuum. 
  This plot represents a subset of the spectral region used in our 
  analysis; the full region is 5\sA\ wide and contains three additional 
  continuum fitting regions.
  \label{eu4129_nso}}
\end{figure}

\begin{figure}  
\includegraphics[width=\columnwidth]{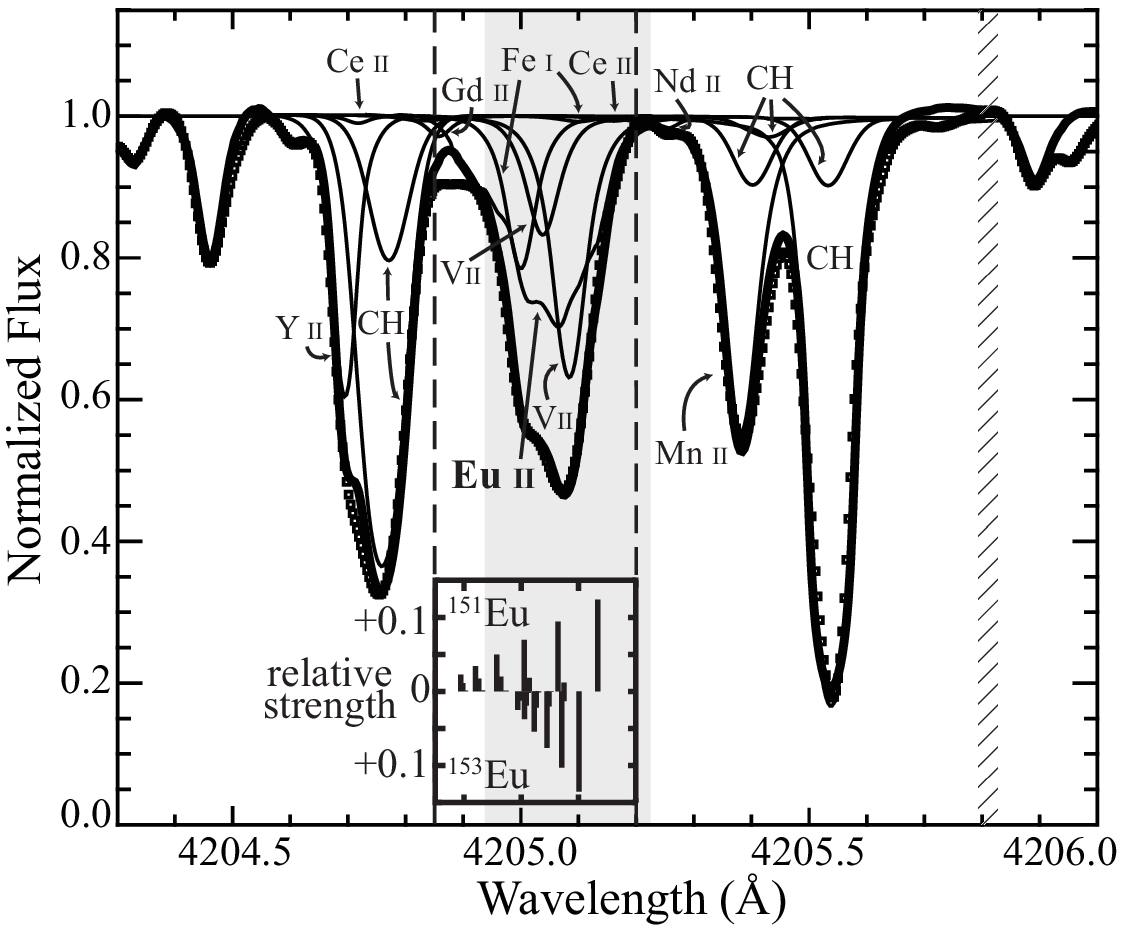}
\figcaption[Eu 4205\sA\ in solar atlas.]{The 4205-\AA\ \euii\ line in the solar spectrum. 
  Individual lines are annotated and are listed in Table \ref{4205_nso_table}. 
  The hyperfine components (see \S\,\ref{linelists}) appear in the inset plot, 
  which is aligned with the wavelength scale of the main plot.
  The relative strengths of the 30 hyperfine components \citep{ivans_2006}
  are plotted on a linear scale in the inset; the top half of the inset 
  contains the components from the $^{151}$Eu isotope while the
  $^{153}$Eu isotope components appear on the bottom.
  The gray box indicates the portion of the spectrum used to calculate 
  \chisq\ during the abundance fitting step (see \S\,\ref{stellar_abund}).  
  The cross-hatched region indicates a portion of the spectrum 
  used to fit a continuum. This plot represents a subset of the 
  spectral region used in our analysis; the full region is 8\sA\ wide 
  and contains six additional continuum fitting regions.
  \label{eu4205_nso}}
\end{figure}

\begin{figure}  
\includegraphics[width=\columnwidth]{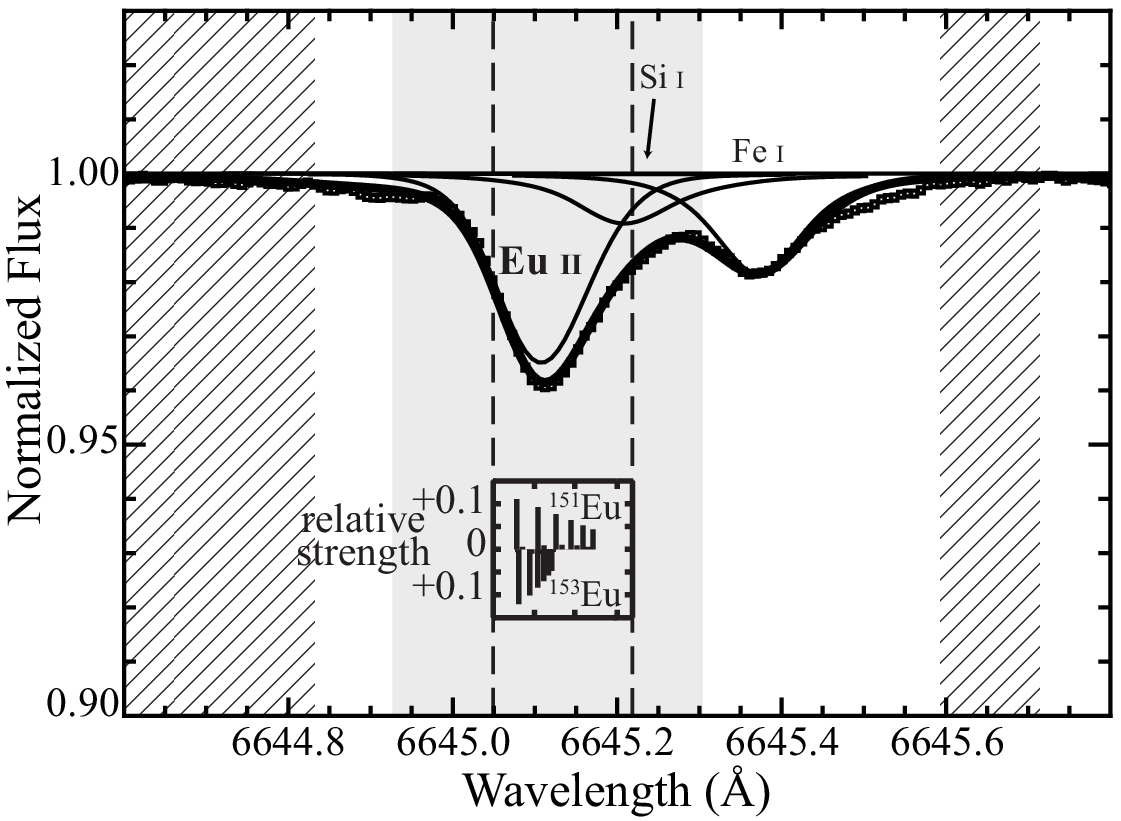}
\figcaption[Eu 6645\sA\ in solar atlas.]{The 6645-\AA\ \euii\ line in the solar 
  spectrum. Note that the ordinate axis is scaled differently than in Figures 
  \ref{eu4129_nso} and \ref{eu4205_nso}.
  Individual lines are annotated and are listed in Table \ref{6645_nso_table}.
  The hyperfine components (see \S\,\ref{linelists}) appear in the inset plot, 
  which is aligned with the wavelength scale of the main plot.
  The relative strengths of the 30 hyperfine components \citep{ivans_2006}
  are plotted on a linear scale in the inset; the top half of the inset 
  contains the components from the $^{151}$Eu isotope while the
  $^{153}$Eu isotope components appear on the bottom.
  The gray box indicates the portion of the spectrum used to calculate 
  \chisq\ during the abundance fitting step (see \S\,\ref{stellar_abund}).  
  The two cross-hatched regions indicate the portion of the spectrum used 
  to fit a continuum.
  This plot represents a subset of the spectral region used in our analysis; 
  the full region is 5\sA\ wide and contains four additional continuum fitting regions.
  \label{eu6645_nso}}
\end{figure}

Where the VALD line list is insufficient, we add line data 
from \citealt{moore_1966}, 
NIST (whose lists are based on a variety of sources) and 
C.~Sneden (private communication).
We also add CH and CN molecular lines based on values
obtained from the Kurucz molecular line list web 
site.\footnote{\tt http://kurucz.harvard.edu/LINELISTS/LINESMOL/}
In the 4129-\AA\ region, we find it necessary to add 
three artificial iron lines in order to match the solar spectrum.
We follow here the precedent of \citealt{delpeloso_2005a},
though we find our fit requires the lines to have 
slightly different wavelengths and \loggf\ values.
The complete line list for \euii\ at 4129\sA\ appears in
Table \ref{4129_nso_table}, 
for 4205\sA\ in Table \ref{4205_nso_table}, and
for 6645\sA\ in Table \ref{6645_nso_table}.
The corresponding plots of the regions in these tables appear in
Figures \ref{eu4129_nso}, \ref{eu4205_nso}, and \ref{eu6645_nso}.

\begin{deluxetable*}{ccccccc}[!htp]
\tablecaption{Line list for the region near \euii\ at 4129\sA.\label{4129_nso_table}}
\tablehead{
  \colhead{$\lambda$} 
  & \multirow{2}{*}{Element}
  & \colhead{Lower Level}
  & \multicolumn{2}{c}{{\loggf}} 
  & \multicolumn{2}{c}{{\gamsix}}\\
  \colhead{(\AA)}
  & 
  & \colhead{(eV)}
  & \colhead{solar fit} 
  & \colhead{VALD\tablenotemark{a}}
  & \colhead{solar fit}
  & \colhead{VALD\tablenotemark{b}}
}

\startdata
   4129.147 &  Pr {\sc ii} &    1.039 & $-0.100$ & $-0.100$ & $-7.454$ &   \ldots \\
   4129.159 &   Cr {\sc i} &    3.013 & $-1.948$ & $-1.948$ & $-6.964$ & $-7.362$ \\
   4129.159 &  Ti {\sc ii} &    1.893 & $-2.300$ & $-1.730$ & $-6.900$ & $-7.908$ \\
   4129.166 &   Ti {\sc i} &    2.318 & $-0.200$ & $-0.231$ & $-6.900$ & $-7.572$ \\
   4129.174 &  Ce {\sc ii} &    0.740 & $-3.000$ & $-0.901$ & $-7.493$ &   \ldots \\
   4129.182\tablenotemark{c} &
                Cr {\sc i} &    2.914 & $-0.100$ &   \ldots & $-8.300$ &   \ldots \\
   4129.220 &   Fe {\sc i} &    3.417 & $-3.500$ & $-2.030$ & $-6.857$ & $-7.255$ \\
   4129.220 &  Sm {\sc ii} &    0.248 & $-1.123$ & $-1.123$ & $-7.536$ &   \ldots \\
   4129.425 &  Dy {\sc ii} &    0.538 & $-0.522$ & $-0.522$ & $-7.554$ &   \ldots \\
   4129.426 &   Nb {\sc i} &    0.086 & $-0.780$ & $-0.780$ & $-7.462$ &   \ldots \\
   4129.458 &   Fe {\sc i} &    3.396 & $-1.950$ & $-1.970$ & $-6.863$ & $-7.206$ \\
   4129.522\tablenotemark{d} &
                Fe {\sc i} &    3.140 & $-3.497$ &   \ldots & $-6.873$ &   \ldots \\
   4129.643 &   Ti {\sc i} &    2.239 & $-1.987$ & $-1.987$ & $-7.529$ & $-7.529$ \\
   4129.705 &  Eu {\sc ii} &    0.000 & $+0.260$ & $+0.173$ & $-7.174$ &   \ldots \\
   4129.817 &   Co {\sc i} &    3.812 & $-1.808$ & $-1.808$ & $-6.099$ & $-7.782$ \\
   4129.837 &  Nd {\sc ii} &    2.024 & $-0.543$ & $-0.543$ & $-6.237$ &   \ldots \\
   4129.959\tablenotemark{d} &
                Fe {\sc i} &    2.670 & $-3.139$ &   \ldots & $-7.322$ &   \ldots \\
   4130.035 &   Fe {\sc i} &    1.557 & $-3.900$ & $-4.345$ & $-7.885$ & $-7.826$ \\
   4130.036 &   Fe {\sc i} &    3.111 & $-2.350$ & $-2.636$ & $-8.026$ & $-7.857$ \\
   4130.073 &   Cr {\sc i} &    2.914 & $-1.971$ & $-1.971$ & $-6.929$ & $-7.349$ \\
   4130.122\tablenotemark{e} &
                V  {\sc i} &    1.218 & $-1.000$ & $-3.142$ & $-7.060$ & $-7.800$ \\ 
   4130.233 &   Mn {\sc i} &    2.920 & $-2.400$ & $-3.309$ & $-7.900$ & $-7.784$ \\
   4130.315 &   V  {\sc i} &    2.269 & $-0.300$ & $-0.607$ & $-7.187$ & $-7.585$ \\ 
   4130.364 &  Gd {\sc ii} &    0.731 & $+0.177$ & $-0.090$ & $-6.608$ &   \ldots \\
   4130.452 &   Cr {\sc i} &    2.914 & $-1.099$ & $-2.751$ & $-6.805$ & $-7.348$ \\
\enddata
\tablenotetext{a}{\linedatafootnote}
\tablenotetext{b}{\vanderwaalsfootnote}
\tablenotetext{c}{\moorefootnote}
\tablenotetext{d}{Artificial iron lines included after \citealt{delpeloso_2005a}.}
\tablenotetext{e}{\kuruczonlinefootnote}
\end{deluxetable*}

Hyperfine splitting is the dominant broadening mechanism for the
europium spectral lines. 
The interaction between the nuclear spin and the atom's 
angular momentum vector causes energy level splitting
in atoms with odd atomic numbers (europium is $Z=63$). 
The effect is particularly pronounced in rare earth 
elements.
The 4129-\AA\ and 4205-\AA\ lines, for example, have FWHMs 
of 1.5\sA, due largely to hyperfine structure
(but not isotope splitting---see insets of Figures 
\ref{eu4129_nso}--\ref{eu6645_nso}).
Since the relative strengths of the hyperfine
components are constant without regard to
temperature, pressure, or magnetic field \citep{abt_1952},
the components measured in laboratory settings
can be applied to stellar spectra.

Like other spectral fitting packages,
SME has no built-in treatment of hyperfine 
structure. 
We therefore convert a single europium line into 
its constituent hyperfine components and include them 
as separate entries in the star's line list.  
The relative strengths and wavelength offsets of the 
hyperfine components come from \citealt{ivans_2006}, 
which bases the values on an FTS laboratory analysis. 
Following the procedure of \citealt{ivans_2006}, 
we assume a solar system 
composition for the relative abundance of the two 
europium isotopes ($^{151}$Eu at 47.8\% and 
$^{153}$Eu at 52.2\%, from \citealt{rosman_1998}).
We divide the \euii\ \loggf\ values listed in 
Tables \ref{4129_nso_table}, \ref{4205_nso_table}, 
and \ref{6645_nso_table} amongst the 
hyperfine components according to their relative strengths.  
All other attributes of the \euii\ lines remain the same in 
the creation of the hyperfine lines.  
The relative strengths of the hyperfine components are displayed 
in the inset plots in the solar spectrum Figures
\ref{eu4129_nso}, \ref{eu4205_nso}, and \ref{eu6645_nso}.

\begin{deluxetable*}{ccccccc}
\tablecolumns{7}
\tablecaption{Line list for the region near \euii\ at 4205\sA.\label{4205_nso_table}}
\tablewidth{0pt}
\scriptsize
\tablehead{
  \colhead{$\lambda$} 
  & \multirow{2}{*}{Element}
  & \colhead{Lower Level}
  & \multicolumn{2}{c}{\loggf}
  & \multicolumn{2}{c}{\gamsix} \\
  \colhead{(\AA)}
  & 
  & \colhead{(eV)}
  & \colhead{solar fit} 
  & \colhead{VALD\tablenotemark{a}} 
  & \colhead{solar fit}
  & \colhead{VALD\tablenotemark{b}} 
}

\startdata
   4204.695 &  Y  {\sc ii} &    0.000 &  $-1.800$ &  $-1.760$ &  $-7.700$ &  \ldots \\
   4204.717 &  Ce {\sc ii} &    0.792 &  $-0.963$ &  $-0.963$ &  $-7.000$ &  \ldots \\
   4204.759\tablenotemark{c} &
                        CH &    0.519 &  $-1.140$ &  $-1.158$ &  $-8.900$ &  \ldots \\ 
   4204.771\tablenotemark{c} &
                        CH &    0.520 &  $-1.900$ &  $-1.135$ &  $-8.500$ &  \ldots \\ 
   4204.801 &  Sm {\sc ii} &    0.378 &  $-1.771$ &  $-1.771$ &  $-6.738$ &  \ldots \\
   4204.831\tablenotemark{c} &
                        CH &    0.520 &  $-3.360$ &  $-3.360$ &  $-7.699$ &  \ldots \\ 
   4204.858 &  Gd {\sc ii} &    0.522 &  $-0.668$ &  $-0.668$ &  $-6.787$ &  \ldots \\
   4204.990 &   Cr {\sc i} &    4.616 &  $-1.457$ &  $-1.457$ &  $-6.658$ &  $-7.852$ \\
   4205.000 &   Fe {\sc i} &    4.220 &  $-1.900$ &  $-2.150$ &  $-7.000$ &  $-7.548$ \\
   4205.038 &  V  {\sc ii} &    1.686 &  $-1.850$ &  $-1.875$ &  $-6.800$ &  $-7.913$ \\
   4205.042 &  Eu {\sc ii} &    0.000 &  $+0.250$ &  $+0.120$ &  $-6.800$ &  \ldots \\
   4205.084 &  V  {\sc ii} &    2.036 &  $-1.100$ &  $-1.300$ &  $-6.900$ &  $-7.956$ \\
   4205.098 &   Fe {\sc i} &    2.559 &  $-4.900$ &  $-4.900$ &  $-6.671$ &  $-7.865$ \\
   4205.107 &   Cr {\sc i} &    4.532 &  $-1.160$ &  $-1.160$ &  $-6.582$ &  $-7.776$ \\
   4205.163 &  Ce {\sc ii} &    1.212 &  $-0.653$ &  $-0.653$ &  $-6.672$ &  \ldots \\
   4205.253 &  Nd {\sc ii} &    0.680 &  $-0.992$ &  $-0.992$ &  $-6.699$ &  \ldots \\
   4205.303 &   Nb {\sc i} &    0.049 &  $-0.850$ &  $-0.850$ &  $-6.677$ &  \ldots \\
   4205.381 &  Mn {\sc ii} &    1.809 &  $-3.300$ &  $-3.376$ &  $-6.800$ &  $-8.001$ \\
   4205.402\tablenotemark{c} &
                        CH &    0.488 &  $-2.300$ &  $-3.960$ &  $-8.000$ &  \ldots \\ 
   4205.427\tablenotemark{c} &
                        CH &    1.019 &  $-2.300$ &  $-1.130$ &  $-8.000$ &  \ldots \\ 
   4205.491\tablenotemark{c} &
                        CH &    1.019 &  $-3.463$ &  $-3.463$ &  $-8.000$ &  \ldots \\ 
   4205.533\tablenotemark{c} &
                        CH &    1.019 &  $-1.800$ &  $-1.149$ &  $-8.000$ &  \ldots \\ 
   4205.538 &   Fe {\sc i} &    3.417 &  $-1.100$ &  $-1.435$ &  $-7.800$ &  $-7.224$ \\
\enddata
\tablenotetext{a}{\linedatafootnote}
\tablenotetext{b}{\vanderwaalsfootnote}
\tablenotetext{c}{\moleculesfootnote}
\end{deluxetable*}

\begin{deluxetable*}{ccccccc}
\tablecolumns{7}
\tablecaption{Line list for the region near \euii\ at 6645\sA.\label{6645_nso_table}}
\tablewidth{0pt}
\scriptsize
\tablehead{
  \colhead{$\lambda$} 
  & \multirow{2}{*}{Element}
  & \colhead{Lower Level}
  & \multicolumn{2}{c}{\loggf}
  & \multicolumn{2}{c}{\gamsix} \\
  \colhead{(\AA)}
  & 
  & \colhead{(eV)}
  & \colhead{solar fit} 
  & \colhead{VALD\tablenotemark{a}} 
  & \colhead{solar fit}
  & \colhead{VALD\tablenotemark{b}} 
}

\startdata
   6644.320\tablenotemark{c} &
                         CN &    0.805 & $-1.456$ & $-2.258$ &  $-7.695$ &  \ldots \\
   6644.415\tablenotemark{d} &
                 La {\sc i} &    0.131 & $-1.330$ & $-2.070$ &  $-8.000$ &  \ldots \\
   6645.111 &  Eu {\sc ii} &    1.380 & $+0.219$ & $+0.205$ &  $-7.218$ &  \ldots \\
   6645.210 &   Si {\sc i} &    6.083 & $-2.510$ & $-2.120$ &  $-7.118$ &  \ldots \\
   6645.372 &   Fe {\sc i} &    4.386 & $-2.759$ & $-3.536$ &  $-6.780$ &  $-7.808$ \\
\enddata
\tablenotetext{a}{\linedatafootnote}
\tablenotetext{b}{\vanderwaalsfootnote}
\tablenotetext{c}{\moleculesfootnote}
\tablenotetext{d}{\kuruczonlinefootnote}
\end{deluxetable*}

\subsection{Europium Abundances}
\label{stellar_abund}
In order to measure the europium abundances in our selected stars, 
we begin with the co-added spectra described in \S\,\ref{data}. 
We do a preliminary continuum fit using the SME routines,
which follow the VF05 procedure:\ deep features are filled 
in with a median value from neighboring spectral orders, 
then a sixth-order polynomial is fit to the region of interest.
The built-in procedure creates a flat, normalized
continuum, though we find it necessary to fine-tune the 
continuum normalization in the course of our europium 
fitting.

From the VF05 SPOCS catalog we take 
\teff, \logg, \monh, $v\sin{i}$, $v_{mac}$, and $v_{mic}$
(fixed at 0.85\unit{\kms}) 
for each star we consider. 
In general, the global parameters from VF05 agree very well with
the values adopted in the studies we compare to here. 
(Most of the literature values fall within the 2-$\sigma$ errors
quoted in VF05.) The VF05 catalog is one of the largest and most 
reliable sources of stellar properties determined to date; 
\citealt{haywood_2006}, for example, finds the VF05 \teff\ and \monh\
to be in good agreement with other reliable measurements.

For most element abundances we use a scaled solar
system composition, shifting the 
\citealt{grevesse_1998} solar abundances by 
the star's [M/H].
The exceptions to this rule are 
sodium, silicon, titanium, iron, and nickel,
which VF05 measured individually. 
Those $\alpha$ and iron-peak elements are
therefore treated independently of the na\"ive scaled-solar
adjustment.
It is possible that a more explicit treatment of iron-peak
and $\alpha$ elements would improve our europium measurement
accuracy. 
For the present study, however, we deem individual 
abundance analysis (apart from europium) unnecessary.
We hold fixed the abundances of all elements other than 
europium in the subsequent analysis.

We fit for the europium abundance by iterating three 
\chisq-minimization routines that solve for the 
wavelength alignment, spectrum continuum, and 
europium abundance.
A summary of each routine follows:
\begin{enumerate}
\item Wavelength. 
The pixel scale is pre-determined from
the thorium lamp calibration taken each night.
A first estimate of the rest frame wavelengths comes from
a cross-correlation of the full spectral segment with the 
solar spectrum, a built-in functionality of SME.
We then use a 2-\AA\ region immediately surrounding 
the \euii\ line of interest to perform a 
\chisq\ minimization between the modeled stellar 
atmosphere and the spectral data, thus solving for
the wavelength scale alignment as precisely as 
possible in the \euii\ region.
\item Continuum.
We fit a quadratic function across the points designated 
in the solar spectrum as continuum-fitting points 
(the cross-hatch regions in 
Figures \ref{eu4129_nso}, \ref{eu4205_nso},
and \ref{eu6645_nso}).
We adjust the quadratic continuum function vertically
to require that 1--2\% of spectral points 
in the full spectral segment are above unity, 
thereby ensuring that all spectra are scaled identically.
\item Abundance.
We perform a \chisq\ minimization 
adjusting only the abundance of europium.
We begin with the solar abundance value scaled
by the star's metallicity, then search 
$1.0\unit{dex}$
of europium abundance space to find the 
best-fit value.
In minimizing the \chisq\ statistic, 
we calculate the residuals between the data 
and the fit in a limited region around the 
\euii\ line (the gray regions in Figures 
\ref{eu4129_nso}, \ref{eu4205_nso}, and 
\ref{eu6645_nso}).
\end{enumerate}
\nin The wavelength alignment, spectrum continuum, and europium abundance 
fitting routines are run in that order and iterated 
until a stable solution is reached.
In most cases a stable solution requires only one or 
two iterations.
The abundances for each line in each star as 
determined by this process are listed in Table 
\ref{starsvals_table}.

\begin{figure}  
\includegraphics[width=\columnwidth]{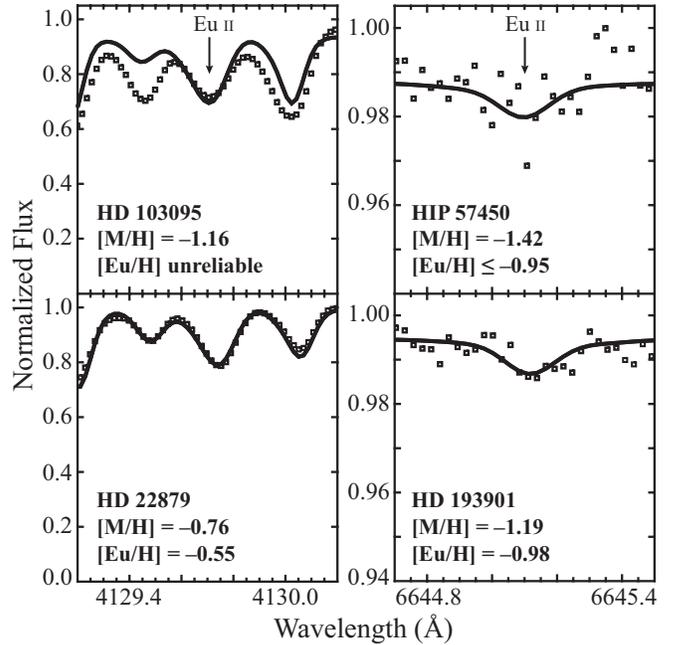}
\figcaption[Bad fits vs.\ good fits in comparable stars.]{Bad fits (top) 
  versus good fits (bottom) in comparable stars. 
  In the upper left panel, the SPOCS stellar properties fit the lines adjacent to
  \euii\ poorly enough that \euh\ is unreliable. That line is removed from further 
  analysis.
  In the upper right panel, the 6645-\AA\ line is buried in the noise, meaning
  the fit represents an upper limit to the europium abundance. The abundance
  upper limit is noted in Table \ref{starsvals_table}, but does not contribute
  to the overall \euh\ measurement in the star.
  The lower two panels include fits to stars with similar characteristics 
  to the stars in the upper panels, but where the 4129-\AA\ and 6645-\AA\
  fits were more successful.
  \label{keep}}
\end{figure}

After running our automatic europium fitting algorithm 
on all \nstarsn\ stars,
we examine each line in each star by eye to 
confirm that the fit is successful. 
In a few of the metal-poor 4129-\AA\ fits, the blended lines
that encroach on \euii\ were poorly enough fit with the
VF05 SPOCS values that the europium value was 
unconvincing. 
In those cases, the 4129-\AA\ value is
omitted from Table \ref{starsvals_table} and does not
contribute to the average.
See Figure \ref{keep} for a comparison of a rejected
4129-\AA\ feature and a robust 4129-\AA\ fit.
Similarly, in a few metal-poor stars the
6645-\AA\ line is too weak to be seen in the noise, 
and hence the output of the fitting routine serves only
as an upper limit on the europium abundance.
See Figure \ref{keep} for a comparison of a measureable 
6645-\AA\ feature and a feature that provides an 
upper limit.
In the cases where the 6645-\AA\ line is an upper limit
only, it is listed as such in Table \ref{starsvals_table}
and does not contribute to the average. 

If both 4129\sA\ and 6645\sA\ proved problematic we removed the
star from our analysis entirely. 
The three stars for which this was the case (listed in
\S\,\ref{stellsamp}) are omitted from 
Tables \ref{starsinfo_table} and \ref{starsvals_table}.
Since all of the rejected stars were
metal poor, we conclude that our 
fitting routine is most robust at solar metallicity, 
and becomes less reliable at \monh~$< -1$. 
Temperature may also play a role, as one of the
rejected stars, HD\,64090, has $\mteff = 7300\unit{K}$; 
VF05 determined SME to be reliable between 4800 and 
$6500\unit{K}$.

For each star we calculate a weighted average europium 
abundance value based on the three (or, in some 
cases, two) spectral lines.
Weighting the average is important because for stars with
relatively few observations (e.g., HD\,156365 in Figure 
\ref{coadding_works}), the weak line at 6645\,\AA\ should 
be weighted significantly less than the more robust blue 
lines. 
Stars with a larger number of observations 
and higher S/N spectra
(e.g., HD\,144585 in Figure \ref{coadding_works})
should have the 6645-\AA\ line weighted more strongly.
In order to determine the relative weights of the three 
spectral lines, we tested the robustness of our fit by 
adding artificial noise to the spectra. 
We describe that process in \S\,\ref{errors}.

\subsection{Error Analysis}
\label{errors}
We begin our error analysis by adding to the data 
Gaussian-distributed random noise with a standard 
deviation set by the photon noise at each pixel.
We then fit the europium line again, using the 
same iterative \chisq-minimization process described 
in \S\,\ref{stellar_abund}, and repeat the process 
50 times.
The standard deviation of the 50 Monte Carlo 
trials determines the relative weights of the 
lines in the average listed in 
Table \ref{starsvals_table}, with lower 
standard deviation lines (corresponding 
to a more robust fit) weighted more 
strongly.

As expected, the results of the Monte Carlo trials show 
that the larger the photon noise in an observation, the 
less that line is to be weighted. 
We derive a linear relation between 
the Poisson uncertainty of an observation
and its relative weight, based on the Monte Carlo results.
Because the Monte Carlo trials are CPU intensive, we plan
to use that relation to determine 
relative weights in the future.

We estimate the uncertainty of our europium
abundances to be the sum of the squares of the 
residuals for all three lines, where 
we assign the residuals the same relative
weights we applied when calculating the average.
We also impose an error floor of $0.02\unit{dex}$,
added to the uncertainty in quadrature.

\begin{figure}  
\includegraphics[width=\columnwidth]{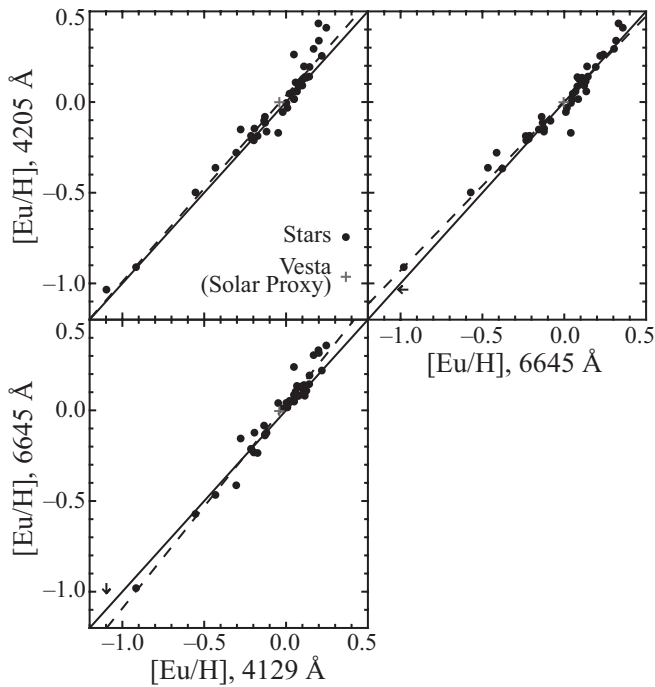}
\figcaption[Comparing three \euii\ lines.]{A comparison of the \euii\ values 
  measured in the \nstarsn\ stars of this study, as listed in Table 
  \ref{starsvals_table}. 
  Vesta is marked with a plus symbol.
  The solid line represents a 1:1 correlation; the dashed line is a 
  best fit to the data.
  Points are omitted where the fit was poor in one of the \euii\ lines. 
  Upper limits for the 6645-\AA\ line are marked with arrows.
  See \S\,\ref{compare_lines} for a discussion of the quality of the fit.
  \label{line_compare}}
\end{figure}

The error floor comes from a comparison of
abundance values from the individual lines
(Figure \ref{line_compare}, discussed in more 
detail in \S\,\ref{compare_lines});
error bars of $0.03\unit{dex}$ on each measurement
would make the \rchisq~$=1$. 
A minimum error of $0.03\unit{dex}$ on each 
europium measurement
translates into an uncertainty of $0.02\unit{dex}$ 
for the final averaged value.
The $0.02\unit{dex}$ error floor is included in the 
uncertainty quoted in Table \ref{starsvals_table}.

As an initial test of the robustness of our fitting routine,
we measure the europium abundance in a spectrum of Vesta,
which serves as a solar proxy.
The values are listed in the last row of
Table \ref{starsvals_table}.
The three Vesta europium abundances have a standard
deviation of $0.022\unit{dex}$, indicating that 
systematic errors ($\sim$\,$0.03\unit{dex}$ for Vesta,
its offset from the solar value) may be a significant
portion of the error budget.

For the moment we absorb any systematic error with 
our random error estimates.
The full sample of 1000 stars, the analysis 
of which will follow this work, will
allow a far more thorough investigation of 
the dependence of our results on various model 
parameters (\teff, \logg, etc.)\ than can be 
accomplished here.
Therefore, we delay a substantive discussion of 
systematics until 
we have more europium abundance measurements in hand,
though we touch on it again in \S\,\ref{compare_lines}
and \S\,\ref{compare_lit}.
It is important to note that since most of the CCPS stars 
are similar to the Sun, and we are treating each star
identically, our results will be internally consistent.

\subsection{Notes on Individual Lines}
\label{individual_lines}

\subsubsection{Europium 4129\sA}
\label{abund:eu4129}
The europium line at 4129\sA\ is the result of a 
resonance transition of \euii, and is a strong, 
relatively clean line.  
It provides the most reliable measurement of 
europium abundance in a star.  
Our fit to the \euii\ line at 4129\sA\ in the 
solar spectrum (described in \S\,\ref{linelists}) 
appears as Figure \ref{eu4129_nso}, with the 32
hyperfine components (16 each from $^{151}$Eu 
and $^{153}$Eu; \citealt{ivans_2006}) 
represented in the Figure 
\ref{eu4129_nso} inset.

In the course of stellar fitting, the europium 
abundance is determined from the 4129\sA\ line 
in all \nstarsn\ stars except HD\,103095. 
That star is quite metal poor (\monh~$= -1.16$) and cool 
($\mteff = 4950$), factors that likely contribute to the 
poor fit in the lines adjacent to the \euii\ line of 
interest (see Figure \ref{keep}).

\subsubsection{Europium 4205\sA}
\label{abund:eu4205}
The \euii\ line at 4205\sA\ is the other fine-structure
component of the resonance transition responsible
for the line at 4129\sA.
Though it is as strong as the \euii\ line at 4129\sA, 
contamination from embedded lines (see Figure \ref{eu4205_nso})
has the potential make the 4205-\AA\ line less reliable.  
However, it is useful as a comparison line for the results
from the 4129-\AA\ \euii\ line.
Our solar spectrum fit at 4205\sA\ appears in 
Figure \ref{eu4205_nso}, with the  
30 hyperfine components (15 each from $^{151}$Eu and $^{153}$Eu;
\citealt{ivans_2006}) 
in the inset.
Despite its blended nature, the fit appears sound
in all \nstarsn\ stars.

\subsubsection{Europium 6645\sA}
\label{abund:eu6645}
The \euii\ line at 6645\sA\ is weaker than the lines in the blue, but
it is also relatively unblended, making it worthwhile to fit wherever 
possible.  
Our solar spectrum fit at 6645\sA\ appears in 
Figure \ref{eu6645_nso},
with the inset plot showing the 30 hyperfine components (15
each from $^{151}$Eu and $^{153}$Eu; \citealt{ivans_2006}). 
In HIP\,57450, which is metal poor (\monh~$= -1.42$) and has only
one observation, the 6645-\AA\ line was lost in the noise
(see Figure \ref{keep}), and
only the 4129-\AA\ and 4205-\AA\ lines contribute to the final
europium abundance; the 6645-\AA\ line provides an upper limit only.

\section{Results}
\label{comp}
\subsection{Comparison of Individual Lines}
\label{compare_lines}
We compare our results from the three \euii\ lines in
Figure \ref{line_compare}.
The Vesta abundances, included as a solar 
proxy, are consistent with the stellar 
abundances to within $0.03\unit{dex}$.
By assigning a measurement uncertainty
of $0.03\unit{dex}$ to each line, the 
\rchisq\ for each of the three plots is unity.
This measurement uncertainty is the source of the
error floor discussed in \S\,\ref{errors}.

\begin{figure*}  
\centering
\includegraphics[width=0.80\textwidth]{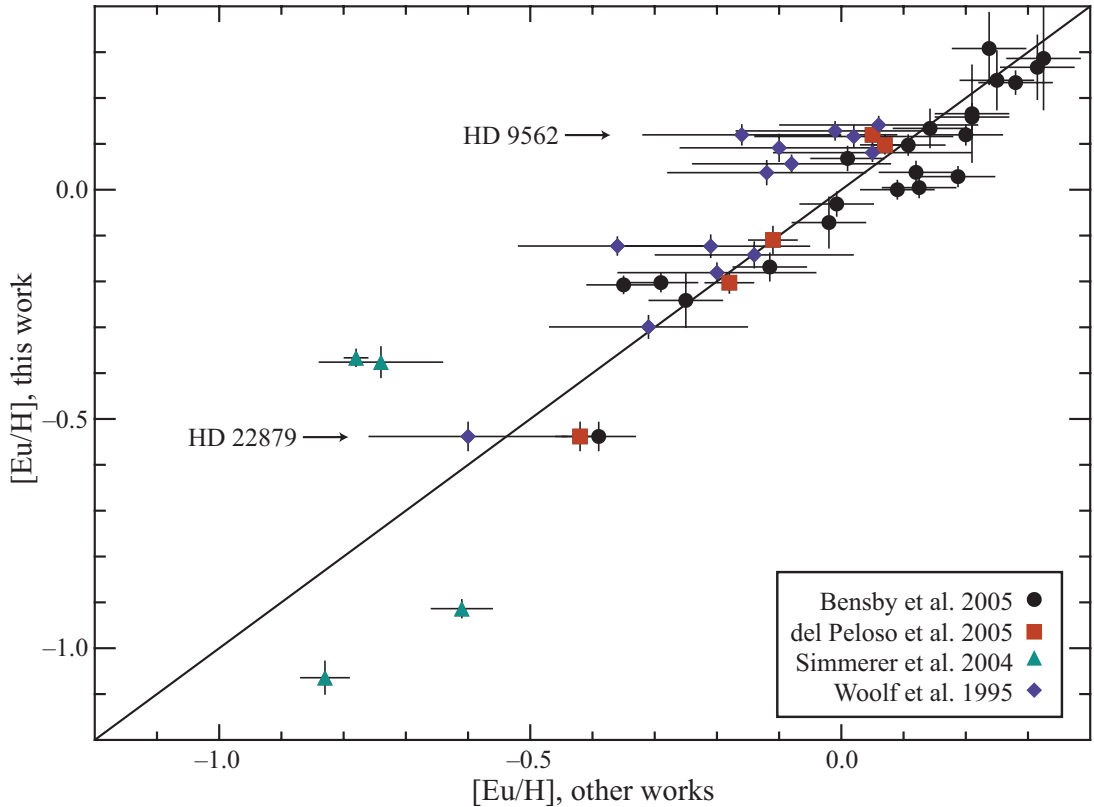}
\figcaption[Comparing our \euii\ values to others'.]{A comparison of the 
  final \euii\ values from this work with literature measurements. 
  The solid line represents a 1:1 correlation. 
  The horizontal error bars represent the uncertainties as quoted in 
  the source literature. 
  The vertical error bars come from our analysis as described in \S\,\ref{errors}. 
  Two stars, HD\,9562 and HD\,22879 (with \euh\ of $+0.12$ and $-0.54$, 
  respectively), were included in the \citealt{bensby_2005}, 
  \citealt{delpeloso_2005a}, and \citealt{woolf_1995} analyses, so those 
  two stars are represented by three points along the abscissa at the 
  same ordinate value. 
  See Table \ref{starsinfo_table} for the full list of the europium 
  abundances plotted here, and see \S\,\ref{compare_lit} for a discussion
  of the quality of our fit based on this plot. 
  \label{me_vs_them}}
\end{figure*}

Calculating the best-fit line for each 
comparison plot (the dashed lines in
Figure \ref{line_compare}) lends insight
into possible systematic trends in our analysis.
In the blue, the two \euii\ lines (4129\sA\ and
4205\sA) have no apparent linear systematic 
trend:\ the slope of the best-fit line is 1.03.
The red \euii\ line (6645\sA), however, exhibits
a minor systematic trend:\ the best-fit line
in the red has a slope of 1.10 relative to the blue,
i.e., a 10\% stretching of the blue abundance values 
about \euh~$= 0$ roughly reproduces the red 
abundance values.
Since this systematic trend is absorbed by the
$0.03\unit{dex}$ error bars assigned to each point 
(necessary to make \rchisq\ unity even when
comparing the non-systematic blue lines),
we make no attempt to correct this systematic
trend here.

Between the relatively low measurement uncertainty
needed to achieve \rchisq\ of unity and the minor
systematic trend that only appears in the red, 
we conclude that the europium abundance values 
derived from the \euii\ lines at 4129\sA, 
4205\sA, and 6645\sA\ are consistent with one 
another.

\subsection{Comparison with Literature Europium Measurements}
\label{compare_lit}
In Figure \ref{me_vs_them}, we compare our 
final europium abundance measurements to the
literature values.
The agreement is quite good.
The literature values, plotted with the error
bars quoted in the original studies, appear
as the abscissa.
Our europium values, plotted with the error
bars calculated in \S\,\ref{errors} and listed
in Table \ref{starsvals_table}, appear
as the ordinate.
The solid line represents a 1:1 correlation.
Comparing the points to the 1:1 line,
the \rchisq~$=\rchisqtot$.
That value is dominated by
HD\,103095, the \citealt{simmerer_2004} data 
point with the smallest error bars, and 
omitting it makes
the \rchisq~$=\rchisqsm$. 

Adopting the global values (\teff, \monh, \logg)
used by the comparison studies (instead of the VF05 
values) drops the \rchisq\ to \rchisqlit, indicating
that some of the scatter in Figure \ref{me_vs_them}
is from the choice of stellar parameters.
We emphasize that the VF05 stellar parameters are 
reliable; we calculate \rchisq\ using
the comparison studies' values in an attempt to 
separate how much of the disagreement in Figure 
\ref{me_vs_them} is from the europium abundance
technique and how much is from the parameters adopted.

Omitting the outlier mentioned at the beginning
of this section, we examine 
the data in Figure \ref{me_vs_them} to 
search for systematics in our results relative 
to the literature values. 
We consider the effects of a global offset in
our europium values and a linear trend
with europium abundance, finding that systematic
offsets of $\sim$\,$0.1\unit{dex}$ and linear trends
of $\sim$\,$40\%$ are needed
to make \rchisq\ climb to 2.
We conclude that our error bars have sufficiently
characterized our uncertainties.

Overall, we find \rchisq\ to be dominated by the 
points at \euh~$< 0.5$, consistent with
our conclusion in \S\,\ref{stellar_abund} that
our results are most reliable near solar 
metallicities, although
the larger error bars on the \citealt{woolf_1995}
points make the correlation very forgiving near
\euh~$=0$. 
Our spectra have very high S/N:\ $\sim$\,160 
at 4200\sA\ 
in a single observation, enhanced substantially
by our co-adding procedure (\S\,\ref{data}).
The automated
SME synthesis treats all stars consistently,
especially important for line blends in the 
crowded blue regions.
For these reasons we believe that our europium abundance
technique is accurate and robust and
that our smaller error bars are warranted.

We conclude that
our abundance measurements are consistent
with previous studies, not surprising
since most abundance techniques rely on the same
Kurucz stellar atmosphere models.
Because the majority of the points in
Figures \ref{me_vs_them} fall near the
1:1 correlation, we also conclude that
near \euh~$=0$ our errors are 
$\sim$\,$0.03\unit{dex}$, though at
\euh~$< -0.5$, the errors may be as high as 
$0.1\unit{dex}$.

\section{Summary}
\label{summary}
We have established that our method for measuring europium in 
solar-metallicity stars using SME is sound.
The resolution and S/N of the Keck HIRES spectra are 
sufficiently high to fit the \euii\ lines in question. 
The values obtained from the 
three europium lines are self-consistent, and our final 
averaged europium value for each of the \nstarsn\ stars in this study
are consistent with the literature values for those stars.

By employing SME to calculate our synthetic spectra,
we are self-consistently modeling all the lines in the regions
of interest. 
Any blending from neighboring lines is treated consistently
from star to star, adding robustness to our europium determination.
Using SME has the added benefit of allowing us to 
adopt the stellar parameters from the SPOCS catalog.
Our automated procedure ensures all stars are treated consistently.

Having established a new method for measuring stellar 
europium abundances, we intend to apply our technique to 
1000 F, G, and K stars from the Keck CCPS survey.
Our analysis of europium in these stars will represent the 
largest and most consistent set of europium measurements in
solar-metallicity stars to date, and will provide
insight into the question of the \rproc\ formation
site and the enrichment history of the Galaxy.

\acknowledgements
The author is indebted to
Geoffrey W.~Marcy, 
Christopher Sneden,
Debra A.~Fischer, 
Jeff A.~Valenti, 
Anna~Frebel,
James W.~Truran,
and Taft E.~Armandroff
for productive and enlightening conversations about the
progress of this work.
Particular thanks are extended to
Christopher Sneden,
Geoffrey W.~Marcy, and
Louis-Benoit Desroches
for their thoughtful comments on this paper.
The author is also grateful to her fellow observers
who collected the Keck HIRES data used
here:\ Geoffrey W.~Marcy, 
Debra A.~Fischer,
Jason T.~Wright, 
John Asher Johnson,
Andrew W.~Howard, 
Chris McCarthy,
Suneet Upadhyay,
R.~Paul Butler,
Steven S.~Vogt,
Eugenio Rivera,
and Joshua Winn.
We gratefully acknowledge the dedication of the
staff at Keck Observatory, particularly Grant Hill
and Scott Dahm for their HIRES support.
This research has made use of 
the SIMBAD database, operated at CDS, Strasbourg, France; 
the Vienna Atomic Line Database; 
the Kurucz Atomic and Molecular Line Databases; 
the NIST Atomic Spectra Database;
and NASA's Astrophysics Data System Bibliographic Services.
The author extends thanks to those of Hawaiian 
ancestry on whose sacred mountain of Mauna Kea we
are privileged to be guests. 
Without their generous hospitality, the Keck observations
presented here would not have been possible.

\bibliographystyle{apj}

\end{document}